\documentclass[aps,prl,superscriptaddress,twocolumn,showpacs,amsmath,amssymb]{revtex4}
\usepackage{graphicx}

\begin{document}

\title{Limitations of the Giant Spin Hamiltonian in Explaining Magnetization Tunneling in a Single-Molecule Magnet}

\author{A. Wilson}
\affiliation{Department of Physics, University of Florida,
Gainesville, FL 32611,USA}
\author{J. Lawrence}
\affiliation{Department of Physics, University of Florida,
Gainesville, FL 32611,USA}
\author{E-C. Yang}
\affiliation{Department of Chemistry and Biochemistry, University of
California at San Diego, La Jolla, CA 92093-0358, USA}
\author{M. Nakano}
\affiliation{Department of Applied Chemistry, Osaka University,
Suita 565-0871, Japan}
\author{D. N. Hendrickson}
\affiliation{Department of Chemistry and Biochemistry, University of
California at San Diego, La Jolla, CA 92093-0358, USA}
\author{S. Hill}
\email[corresponding author, Email:]{hill@phys.ufl.edu}
\affiliation{Department of Physics, University of Florida,
Gainesville, FL 32611,USA}

\date{\today}

\begin{abstract}
EPR studies of a Ni$_4$ single-molecule magnet yield the
zero-field-splitting (zfs) parameters, $D$, $B_4^0$ and $B_4^4$,
based on a giant spin approximation (GSA) with $S = 4$. Experiments
on an isostructural Ni-doped Zn$_4$ crystal establish the Ni$^{\rm
II}$ ion zfs parameters. The $4^{\rm th}$-order zfs parameters in
the GSA arise from the interplay between the Heisenberg interaction,
$J\hat{s}_1\cdot\hat{s}_2$, and the $2^{\rm nd}$-order single-ion
anisotropy, giving rise to mixing of higher lying $S\neq4$ states
into the $S = 4$ state. Consequently, $J$ directly influences the
zfs in the ground state, enabling its direct determination by EPR.
\end{abstract}


\maketitle


The [Ni(hmp)($R$OH)Cl]$_4$ molecule (abbreviated Ni$_4$) possessing
the $R$OH~=~dmb ligand (Ni$_4^{\rm dmb}$
\cite{EdwardsJAP03,KirmanJAP,YangPoly03,YangIC05,YangIC06})
represents a model system for carefully examining the validity of
the giant spin approximation (GSA) which has been widely applied in
the study of single-molecule magnets (SMMs) \cite{GatSesAngew}. The
GSA assumes the total spin, {\em S}, of the molecule to be a good
quantum number, and then models the lowest-lying $(2S+1)$ magnetic
sub-levels in terms of an effective spin Hamiltonian of the form:

\begin{equation}\label{eqn:1}
\hat H = D\hat S_z^2 + B_4^0\hat{O}_4^0 + B_4^4\hat{O}_4^4 + \mu _B
\vec B \cdot
\mathord{\buildrel{\lower3pt\hbox{$\scriptscriptstyle\leftrightarrow$}}
\over g} \cdot \hat S.
\end{equation}

\smallskip

\noindent{The first three terms parameterize anisotropic magnetic
interactions which lead to zero-field-splitting (zfs) of the
ground-state multiplet (see red lines in Fig.~1 for the case of
$S=4$), e.g. spin-orbit coupling, dipolar interactions, etc; here,
we consider only $2^{\rm nd}$ and $4^{\rm th}$-order operators (see
\cite{GatSesAngew} for definitions) which are compatible with the
$S_4$ symmetry of the Ni$_4^{\rm dmb}$ SMM. The final term
represents the Zeeman interaction associated with the application of
a magnetic field, $B$, where
$\mathord{\buildrel{\lower3pt\hbox{$\scriptscriptstyle\leftrightarrow$}}
\over g}$ is the Land${\rm \acute{e}}~g$-tensor.}

SMMs are defined by a dominant $2^{\rm nd}$-order uniaxial
anisotropy, $D\hat{S}_z^2$, with $D<0$ \cite{GatSesAngew}.
Nevertheless, weaker $4^{\rm th}$-order terms have been shown to
play a crucial role in the quantum dynamics of several high-symmetry
SMMs (especially
Mn$_{12}$-acetate)~\cite{BokaPRL,BarcoPRL04,OurJLTP,YangIC06}, even
though the precise origin of these terms has not previously been
understood~\cite{Park}. In this letter, we show that higher order
terms [$O(2n$), $n>1$] arise naturally in the GSA through the
interplay between intrinsic magneto-anisotropy (at the sites of
individual magnetic ions in the molecule) and inter-spin-state
mixing (controlled by exchange). These findings raise questions
concerning the validity of the GSA, particularly in terms of its
predictive powers.

\begin{figure}[t]
\centering
\includegraphics*[width=0.85\columnwidth]{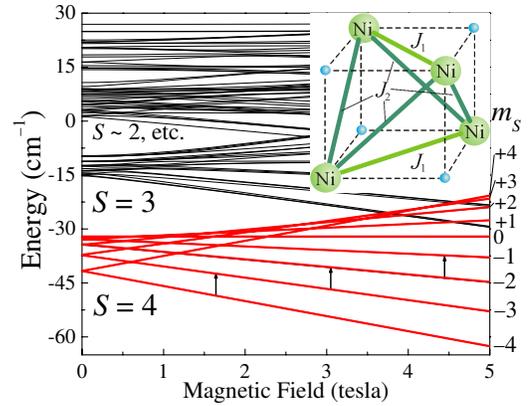}
\caption[]{Field dependence of the 81 eigenvalues corresponding to
the four-spin Hamiltonian (Eq.~\ref{eqn:2}). The lowest 9 levels
(red lines) can be modeled by the GSA with $S=4$ (Eq.~\ref{eqn:1}).
The inset shows a schematic of the cubic core of the Ni$_4^{\rm
dmb}$ SMM (the blue spheres represent O).} \label{fig:1}
\end{figure}

The Ni$_4^{\rm dmb}$ SMM is particularly attractive for this
investigation. The four $s=1$ Ni$^{\rm II}$ ions reside on opposing
corners of a slightly distorted cube (Fig.~1
inset)~\cite{YangPoly03,YangIC05,YangIC06}. DC susceptibility data
($\chi_{\rm M}$T) indicate a relatively large ground state spin of
$S=4$ for the molecule, and a reasonable separation ($\sim35$~K)
between this and higher lying states with
$S<4$~\cite{YangPoly03,YangIC06}. These properties can be
rationalized in terms of pure ferromagnetic coupling between the
Ni$^{\rm II}$ ions. In addition, efforts to fit low-temperature
electron paramagnetic resonance (EPR) and magnetization data to the
GSA (Eq.~\ref{eqn:1} and red lines in Fig.~\ref{fig:1}) have been
highly successful~\cite{EdwardsJAP03,KirmanJAP,YangPoly03}. Thus,
Ni$_4^{\rm dmb}$ displays all of the hallmarks of a SMM, yet it
exhibits unusually fast magnetic quantum tunneling (MQT) at zero
field~\cite{YangIC06}.

The GSA assumes $S$ to be rigid, thereby ignoring the internal
magnetic degrees of freedom within a SMM which can give rise to
couplings to higher-lying states ($S-$mixing
\cite{CarrettaPRL04,HillScience}) that may ultimately influence MQT.
A more physical model, which takes into account zfs interactions at
the individual Ni$^{\rm II}$ sites, as well as the exchange coupling
between individual magnetic ions, is given by the following
Hamiltonian \cite{CarrettaPRL04}:

\begin{equation}\label{eqn:2}
\begin{array}{l}
\hat H = \sum\limits_i {\sum\limits_{j > i} J_{ij} \hat s_i \cdot
\hat s_j } \\ \ \ \ \  + \sum\limits_i {\left[ {d_i\hat s_{zi}^2 +
e_i\left( {\hat s_{xi}^2  - \hat s_{yi}^2 } \right) + \mu _B \vec B
\cdot
\mathord{\buildrel{\lower3pt\hbox{$\scriptscriptstyle\leftrightarrow$}}
\over g_i} \cdot \hat s_i } \right]}.
\end{array}
\end{equation}

\noindent{Here, the $\hat{s}_{\alpha i}$ ($\alpha = x, y, z$)
represent spin projection operators, and $\hat{s}_i$ the total spin
operator, corresponding to the individual Ni$^{\rm II}$ ions; $d_i$
($<0$) and $e_i$ respectively parameterize the uniaxial and rhombic
zfs interactions in the local coordinate frame of each Ni$^{\rm II}$
ion; likewise,
$\mathord{\buildrel{\lower3pt\hbox{$\scriptscriptstyle\leftrightarrow$}}
\over g_i}$ represents the Land${\rm \acute{e}}$ $g$-tensor at each
site; finally, the $J_{ij}$ parameterize the isotropic exchange
couplings between pairs of Ni$^{\rm II}$ ions.}

For Ni$_4^{\rm dmb}$, the dimension of the four-spin Hamiltonian
matrix (Eq.~\ref{eqn:2}) is just $[(2s+1)^4]^2=81\times81$, which is
easily handled by any modern PC (in contrast to Mn$_{12}$-acetate
which has dimension $\sim 10^8\times10^8$~\cite{RaghuPRB01}). More
importantly, the $3\times3$ Hamiltonian matrix associated with a
single Ni$^{\rm II}$ ion contains only two zfs parameters, $d_i$ and
$e_i$ (in addition to
$\mathord{\buildrel{\lower3pt\hbox{$\scriptscriptstyle\leftrightarrow$}}
\over g_i}$). Furthermore, due to the high symmetry of the molecule,
these matrices are related simply by the $S_4$ symmetry operation,
and the number of exchange constants reduces to just two ($J_1$ and
$J_2$, see Fig.~\ref{fig:1} inset). Consequently, the four-spin
model contains only a hand full of parameters, each of which can be
determined independently, often by more than one
method~\cite{EdwardsJAP03,KirmanJAP,YangPoly03,YangIC05,YangIC06}.
Fig.~\ref{fig:1} displays the 81 Zeeman-split eigenvalues
corresponding to the four-spin Hamiltonian (Eq.~\ref{eqn:2}), using
parameters obtained from fits described later. The lowest nine
levels are fairly well separated from higher lying states; these
levels, which dominate the EPR spectrum, can be equally well
described by the Hamiltonian corresponding to Eq.~\ref{eqn:1} with
$S=4$~\cite{EdwardsJAP03,KirmanJAP,YangPoly03}. Roughly $20~{\rm
cm}^{-1}$ above this ground state multiplet exists a grouping of 21
levels which can reasonably be treated as three separate $S=3$
multiplets. There is then a gap to a more-or-less continuum of
levels. The notion of a well defined spin quantum number becomes
tenuous at this point.



There are a number of other important reasons why we chose to focus
on the Ni$_4^{\rm dmb}$ member of the Ni$_4$ family. To begin with,
Ni$^{\rm II}$ is readily amenable to substitution with non-magnetic
Zn. Thus, one can synthesize crystals of Zn$_4^{\rm dmb}$ lightly
doped with Ni$^{\rm II}$~\cite{YangIC05}. The result is a small
fraction of predominantly $s=1$ Zn$_3$Ni magnetic species diluted
into a non-magnetic host crystal. X-ray studies indicate that the
structures of the Ni$_4^{\rm dmb}$ and Zn$_4^{\rm dmb}$ complexes
are virtually identical. Thus, EPR studies of the doped crystals
provide very reliable estimates of the single-ion zfs parameters for
Ni$^{\rm II}$ in the parent Ni$_4^{\rm dmb}$ compound ($d_i$, $e_i$
and
$\mathord{\buildrel{\lower3pt\hbox{$\scriptscriptstyle\leftrightarrow$}}
\over g_i}$ in Eq.~\ref{eqn:2})~\cite{YangIC05}. Another remarkable
feature of the Ni$_4^{\rm dmb}$ member of the Ni$_4$ family is that
its structure contains absolutely no solvent of
crystallization~\cite{YangPoly03,YangIC05,YangIC06}. This is quite
rare among SMMs, resulting in the removal of a major source of
disorder. Indeed, we believe that this is the primary reason why the
Ni$_4^{\rm dmb}$ complex gives particularly sharp EPR
spectra~\cite{JLTPRev,ChakovBrAc}. In contrast, all of the other
solvent containing Ni$_4$ complexes exhibit rather broad EPR
peaks~\cite{EdwardsJAP03}. Details of the experimental procedures,
including representative EPR spectra, are presented
elsewhere~\cite{KirmanJAP,YangIC05,Wilson}.


\begin{figure}[t]
\includegraphics*[width=1\columnwidth]{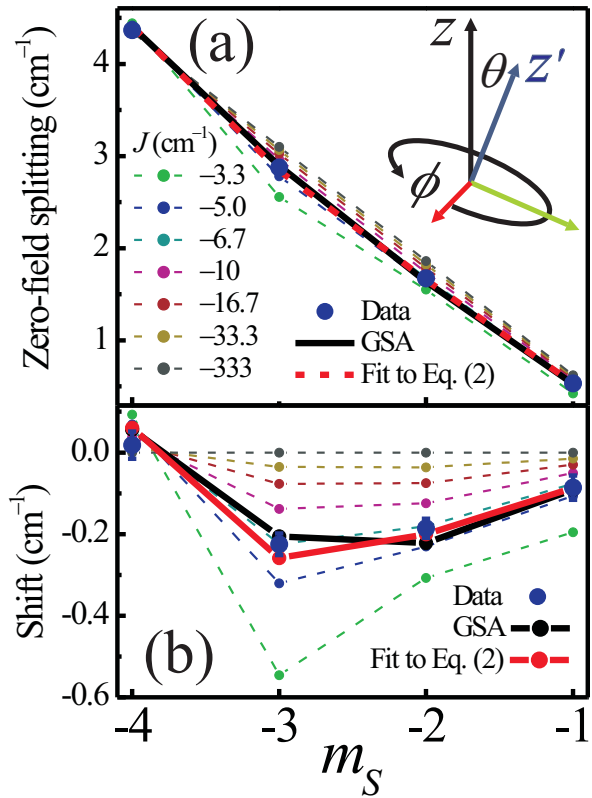}
\caption[]{(a) $m_S$ dependence of the zfs energies [between $m_S$
and $(m_S + 1$)] within the ground state multiplet. The dashed
curves show the zfs obtained from Eq.~\ref{eqn:2} as a function of
$J$. The inset defines the Euler angles relating the Ni$^{\rm II}$
and molecular coordinates~\cite{YangIC05}. (b) Difference between
the data in (a) and the $J=-333~{\rm cm}^{-1}$ curve, emphasizing
the non-linear $m_S$ dependence of the zfs energies.} \label{fig:2}
\end{figure}

We begin by reviewing the results of single-crystal high-frequency
EPR studies of Ni$_4^{\rm
dmb}$~\cite{EdwardsJAP03,KirmanJAP,YangPoly03}. Based on an analysis
using the GSA (Eq.~\ref{eqn:1}), the lowest-lying $S=4$ multiplet is
split by a dominant axial zfs interaction with
$D=-0.589(2)$~cm$^{-1}$. In the absence of higher-order terms, this
interaction produces a quadratic dependence of the ($2S+1$)
zero-field eigenvalues on the quantum number $m_S$, representing the
projection of the total spin onto the easy-axis of the molecule.
Consequently, the zfs between successive $m_S$ levels should be
linear in $m_S$. It is these splittings that one measures in an EPR
experiment, albeit in a finite magnetic field. However, using a
multi-frequency approach, one can extrapolate easy-axis data
($B\parallel z$) to zero-field, yielding accurate determinations of
these splittings~\cite{EdwardsJAP03}; these energy spacings are
plotted versus $m_S$ in Fig.~\ref{fig:2} for Ni$_4^{\rm dmb}$. As
can be seen, the dependence of the zfs values on $m_S$ {\em is not}
linear. One can obtain agreement to within experimental error by
including the 4$^{\rm th}$-order axial zfs interaction
$B_4^0\hat{O}_4^0$ ($\propto\hat{S}_z^4$) in the GSA, with
$B_4^0=-1.2\times10^{-4}$~cm$^{-1}$ (black data in
Fig.~\ref{fig:2}~\cite{EdwardsJAP03}). The $\hat{S}_z^4$ operator
produces quartic $m_S$ corrections to the zero-field eigenvalues
and, thus, cubic corrections to the zfs, as seen in the figure.
Unlike the 2$^{\rm nd}$-order term, which can easily be understood
as originating from the 2$^{\rm nd}$-order zfs interactions at the
individual Ni$^{\rm II}$ sites, the 4$^{\rm th}$-order term in the
GSA does not have any obvious physical meaning. In this sense, it is
nothing more than an adjustable parameter in an effective model
(Eq.~\ref{eqn:1}). As we will see below, this non-linear $m_S$
dependence of the zfs values is directly related to
$S$-mixing~\cite{CarrettaPRL04}.


\begin{figure}[t]
\includegraphics*[width=1\columnwidth]{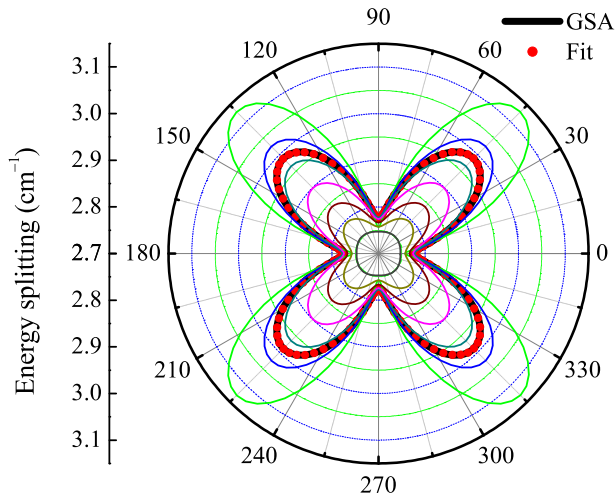}
\caption[]{Angle-dependence of the splitting of the lowest energy
doublet ($m_S=\pm 4$ in zero-field) as a function of the field
orientation within the hard plane. The thin curves are simulations
using Eq.~\ref{eqn:2} with different $J$ values (see
Fig.~\ref{fig:2} for color codes). The black and red data are the
best fits to Eqns.~\ref{eqn:1} and \ref{eqn:2}, respectively.}
\label{fig:3}
\end{figure}

The four-fold ($S_4$) symmetry of the Ni$_4^{\rm dmb}$ molecule
forbids 2$^{\rm nd}$-order zfs interactions which break axial
symmetry. Indeed, we find no evidence for such interactions based on
EPR experiments conducted as a function of the field orientation
within the hard plane. However, a very pronounced four-fold
modulation of the spectrum is observed, which can be explained by
the 4$^{\rm th}$-order $B_4^4\hat{O}_4^4$ [$\equiv
\frac{1}{2}B_4^4(\hat{S}_x^4 + \hat{S}_y^4)$] term in the GSA, with
$B_4^4=\pm 4\times10^{-4}$~cm$^{-1}$~\cite{KirmanJAP}. Although this
interaction is allowed by symmetry, it is not obvious how it relates
to the underlying anisotropy associated with the individual Ni$^{\rm
II}$ ions. Nevertheless, it does explain the fast MQT observed in
this and other Ni$_4$ complexes~\cite{KirmanJAP,YangIC06}. When
treated as a perturbation to the axial zfs Hamiltonian,
$(\hat{S}_x^4+\hat{S}_y^4)$ connects states that differ in $m_S$ by
$\pm4$ in first order and, therefore, lifts the degeneracy of the
lowest lying $m_S=\pm 4$ states in 2nd order, leading to a tunnel
splitting of order 10~MHz. This is an extremely large intrinsic
tunnel splitting in comparison to other SMMs, and can be understood
as arising because of the coincidence of the multiplicity of the
ground state ($2S+1=9$) and the four-fold symmetry, which gives rise
to a leading-order off-diagonal zfs interaction which is fourth
order in the spin operators, i.e. $(\hat{S}_x^4+\hat{S}_y^4)$ is
extremely effective at connecting the $m_S=\pm 4$ states.


We now attempt to understand the physical basis for the existence of
the axial and transverse 4$^{\rm th}$-order zfs interactions ($B_4^0
\hat{O}_4^0$ and $B_4^4 \hat{O}_4^4$) deduced on the basis of the
GSA. From previous studies of a Ni doped Zn$_4^{\rm dmb}$ crystal,
we determined not only the zfs parameters associated with the
Ni$^{\rm II}$ ions, but also the orientations of the local magnetic
axes associated with these interactions relative to the
crystallographic axes~\cite{YangIC05}. However, the key point is
that the Hamiltonian matrices for the individual $s=1$ Ni$^{\rm II}$
ions have dimensions $3\times3$. Therefore, terms exceeding 2$^{\rm
nd}$-order in the single-spin operators ($\hat{s}_{ix}^2$,
$\hat{s}_{iy}^2$, etc.) are completely unphysical. If one assumes
that the ground state for the Ni$_4^{\rm dmb}$ molecule corresponds
to a rigid $S=4$ spin, one can then project the single-ion
anisotropies onto the $S=4$ state using irreducible tensor operator
methods~\cite{YangIC05}. However, after rotating from local to
molecular coordinates, the projection is nothing more than a
summation of the individual zfs matrices. Consequently, such a
procedure {\em does not} produce terms of order four in the spin
operators~\cite{YangIC05}. Therefore, the need to include 4$^{\rm
th}$-order zfs interactions in an analysis of the EPR data for
Ni$_4^{\rm dmb}$ may be taken as evidence for a breakdown of the
GSA. We note that agreement in terms of the 2$^{\rm nd}$-order
parameters is very good. In particular, the molecular $D$ value
agrees to within $10\%$ with the value obtained from projection of
the single-ion anisotropies onto the $S=4$ state~\cite{YangIC05}. In
addition, although the single-ions experience a significant rhombic
zfs interaction ($e/d\sim0.23$), symmetry considerations guarantee
its cancelation when projected onto the $S=4$ state. Therefore, this
approach is completely unable to account for the MQT in Ni$_4^{\rm
dmb}$.

In view of the above, one is forced to use a more realistic
Hamiltonian (Eq.~2) which takes into account all spin states of the
molecule. The isotropic exchange interaction, $J_{ij}$, in Eq.~2
connects states having the same spin-projection~\cite{HillScience}.
Consequently, it does not operate between states within a given spin
multiplet, it simply lifts degeneracies between states with
different multiplicity (see Fig.~\ref{fig:1}). The addition of
anisotropic terms to Eq.~2 results in zfs within each multiplet
which, in turn, gives rise to weaker $m_S$-dependent corrections to
the exchange splittings. Thus, we see that $J$ directly modifies the
zfs energies within a given spin multiplet via interactions
($S$-mixing) with nearby excited spin states. In the limit $J>>d$
one can expect these corrections to be negligible. However, in the
present case, where $J/d\sim1.3$, one can expect these corrections
to be significant. Furthermore, since the corrections involve higher
order processes whereby the underlying anisotropic interactions
feedback into themselves via exchange coupling to nearby
spin-multiplets, it is clear that this will generate `effective'
interactions that are 4$^{\rm th}$-order (i.e. 2$^{\rm nd}$-order
squared) in the spin operators (as well as higher order terms).
However, these 4$^{\rm th}$-order interactions have no real physical
basis other than that they arise due to the competing isotropic and
anisotropic interactions in Eq.~2, resulting in $S$-mixing.

The influence of $J$ on the lowest lying (nominally $S=4$) multiplet
is abundantly apparent in Fig.~2, where we compare zfs energies
determined via the four-spin Hamiltonian (Eq.~\ref{eqn:2}) for
different values of the exchange interaction strength, with those
determined experimentally (blue data points) and from a fit to the
experimental data using the GSA (Eq.~\ref{eqn:1}, black data
points). The magnitudes of $d = -4.73$~cm$^{-1}$ and $e =
-1.19$~cm$^{-1}$ were established from combined fits to both
easy-axis zfs data (red points in Fig.~\ref{fig:2}), and from
hard-plane rotation measurements of the four-fold oscillation of the
ground state splitting (Fig.~\ref{fig:3}, see
also~\cite{KirmanJAP}). We made one simplifying assumption by
setting $J_1=J_2=J$, based on DC $\chi_M T$ data~\cite{Nakano}.
Regardless, this in no way invalidates the main conclusion of this
letter: namely, that $J$ influences the ground state zfs through
$S$-mixing. The polar angle, $\theta$ (see Fig.~\ref{fig:2} inset),
between the local Ni$^{\rm II}$-ion $z$-axes and the
crystallographic $z$-axis was fixed at $15^\circ$ on the basis of
the Ni/Zn studies~\cite{YangIC05}. We additionally included a
dipolar coupling (not shown in Eq.~\ref{eqn:2}) between the four
Ni$^{\rm II}$ ions using precise crystallographic data and no
additional free parameters~\cite{CarrettaPRL04}. The remaining free
parameters were $g_x=g_y=2.23$, $g_z=2.25$ and an additional Euler
angle ($\phi=59^\circ$) illustrated in the inset to
Fig.~\ref{fig:2}. A more in-depth account of the fitting procedure
will be given elsewhere~\cite{Wilson}. The obtained value of $d$
agrees to within $12\%$ with the value determined independently from
measurements on the Ni-doped Zn$_4$ crystal [$d =
-5.30(5)$~cm$^{-1}$]; the remaining parameters agree to within the
experimental error [$e = -1.20(2)$~cm$^{-1}$, $g_x=g_y=2.20(5)$,
$g_z=2.30(5)$].

One can clearly see that, by reducing the separation between the
ground $S=4$ multiplet and the lowest excited states (by reducing
$J$), one can reproduce both the nonlinear $m_S$ dependence of the
zfs energies (Fig.~\ref{fig:2}), which was attributed to the $B_4^0$
term in the GSA~\cite{EdwardsJAP03}, and the four-fold oscillation
of the ground-state splitting observed from hard-plane measurements
(Fig.~\ref{fig:3}), which was attributed to
$B_4^4$~\cite{KirmanJAP}. This is quite a remarkable result, because
it implies that one can deduce $J$ directly from the spectroscopic
information obtained via an EPR experiment. Indeed, the value of
$J=-5.9$~cm$^{-1}$ determined from these fits is in good agreement
with the value of $-7.05$~cm$^{-1}$ deduced on the basis of fits to
$\chi_M T$ data to Eq.~\ref{eqn:2}~\cite{Nakano}. All of the
apparent 4$^{\rm th}$-order behavior vanishes if one sets $J>>d$, as
expected in such a limit in which the ground state spin value is a
good quantum number (due to the absence of $S$-mixing). In the
opposite extreme ($J\sim -3$~cm$^{-1}$), we start to see evidence
for even higher order corrections to the zfs energies (6$^{\rm th}$
order). A cubic polynomial exhibits only one turning point (at
$m_S=-0.5$ in Fig.~\ref{fig:2}), whereas the green data in
Fig.~\ref{fig:2} clearly display more than one turning point when
one recognizes that all of these curves must be antisymmetric about
$m_S=-0.5$. Therefore, it is apparent that one should not limit the
GSA to 4$^{\rm th}$-order terms for SMMs with relatively low-lying
excited spin states. In fact, one cannot rule out equally good fits
to experimental data which include 6$^{\rm th}$ and higher-order zfs
interactions. Consequently, one should be careful about making
predictions on the basis of the GSA, particularly at vastly
different energy scales compared to the experiments used to
establish the GSA zfs parameters (e.g. EPR vs. MQT). Indeed, we find
a difference of almost a factor of 10 between the ground-state
tunnel splittings deduced from Eqs.~1 and 2 using the optimum zfs
parameters for Ni$_4^{\rm dmb}$. We note that the situation in
Ni$_4$ is not dissimilar to many other SMMs, including the most
widely studied Mn$_{12}$-acetate, for which similar 4$^{\rm
th}$-order zfs interactions and low-lying excited spin states are
found~\cite{PetukhovPRB04,CarrettaPRB06}.

Finally, we note that the most unambiguous method for estimating
exchange couplings in polynuclear metal complexes involves
determining the exact locations of excited spin multiplets. However,
the magnetic-dipole selection rule forbids transitions between
states with different multiplicity. Therefore, such an undertaking
is usually only possible using neutrons~\cite{CarrettaPRB06}.
However, Figs.~2 and 3 clearly show that $J$ can be estimated on the
basis of zfs of the lowest lying multiplet. Due to the resultant
$S$-mixing, it may be feasible to observe inter-spin-state EPR
transitions directly via far-infrared techniques.

This work is supported by the National Science Foundation,
DMR-0239481 and DMR-0506946.




\bigskip






\end{document}